\documentclass[aps,prb,superscriptaddress,reprint,amsmath,amssymb]{revtex4-1}
\usepackage{graphicx}
\usepackage{epsf}
\usepackage{bm}
\usepackage{here}
\usepackage{color}
\def\gs{\gtrsim}
\def\ls{\lesssim}

\begin{document}
\title{
Identifying time scales for violation/preservation of Stokes--Einstein relation in supercooled water
}

\author{Takeshi Kawasaki}
\affiliation{
Department of Physics, Nagoya University, Nagoya 464-8602, Japan
}

\author{Kang Kim}
\affiliation{
Division of Chemical Engineering, Graduate School of Engineering
Science, Osaka University, Osaka 560-8531, Japan
}

\date{\today}

\begin{abstract}
The violation of Stokes--Einstein (SE) relation $D\sim (\eta/T)^{-1}$
 between the shear viscosity $\eta$
and the translational diffusion constant $D$ at temperature $T$ is of great importance for
 characterizing anomalous dynamics of supercooled water.
Determining which time scales play key roles in the SE violation remains elusive without the measurement of $\eta$.
Here we provide comprehensive simulation results of the dynamic
properties involving $\eta$ and $D$ in TIP4P/2005 supercooled water.
This enabled the thorough identification of the appropriate time scales
 for SE relation $D\eta/T$.
In particular, it is demonstrated that the temperature dependence of various 
time scales associated with structural relaxation, hydrogen bond breakage, stress
 relaxation, and dynamic heterogeneities
can be definitely classified into only two classes.
That is, we propose the generalized SE relations that exhibit
``violation'' or ``preservation.''
The classification depends on the examined time
scales that are coupled or decoupled with the diffusion.
On the basis of the classification, we explain the physical origins of
 the violation in terms of the increase in the plateau modulus and
the nonexponentiality of stress relaxation.
This implies that the mechanism of SE violation is attributed to the
attained solidity upon supercooling, which is in accord with the growth of
non-Gaussianity and spatially heterogeneous dynamics.
\end{abstract}

\maketitle

\section*{Introduction}

For simple liquids, the Stokes--Einstein (SE) relation between the shear viscosity $\eta$
and the translational diffusion constant $D$ is an
important characteristic of their transport properties~\cite{Hansen:2013uv}.
Specifically, this relation implies $D\sim (\eta / T)^{-1}$, where
$T$ is the temperature.
However, when liquids are supercooled below their melting temperatures,
the SE relation is remarkably violated (SE violation), particularly near the glass
transition temperature~\cite{Hodgdon:1993ew, Stillinger:1995fu,
Tarjus:1995gx, Cicerone:1995dh, Ediger:2000ed, Shi:2013ji,
Sengupta:2013dg, Henritzi:2015jpa}.
Despite extensive efforts, the origins of
SE violation in supercooled liquids remain elusive.

Generally, transport coefficients such as $D$ and $\eta$ are mostly coupled
at high temperatures.
The characteristic time scale is associated with the structural
$\alpha$-relaxation time $\tau_{\alpha}$.
By contrast, at supercooled states, the SE violation implies that
$D$ and $\eta$ are determined by different time scales.
Structural relaxations in supercooled liquids
become spatially heterogeneous, which is a 
different behavior than the homogeneous dynamics observed in normal
liquids~\cite{Ediger:2000ed, Debenedetti:2001bh, Berthier:2011wv}.
Thus, the physical implication of SE violation is relevant to the
question regarding
which time scales determine the transport coefficients in glass-forming liquids.
Alternative types of the SE relation
$D\sim {\tau_\alpha}$ or $D\sim {\tau_\alpha}/T$ have
been controversially tested by assuming that 
$\tau_\alpha$ is proportional to $\eta/T$ (analogous to the
Gaussian approximation) or $\eta$ (analogous to the Maxwell model),
respectively~\cite{Shi:2013ji, Sengupta:2013dg}.

For liquid water, various anomalies in both its thermodynamics and
dynamics have been observed upon supercooling~\cite{Angell:1983bw,
Debenedetti:2003gd, Debenedetti:2003gn, Stanley:2007gf, Gallo:2016fd}.
The SE violation is one of the important anomalies that has been widely reported for
supercooled water~\cite{Chen:2006kk, Becker:2006ju, Kumar:2007hl, Mazza:2007kr,
Xu:2009hq, Chen:2009hx, Mallamace:2010bga, Jana:2011fj, Dehaoui:2015ii}.
In the previous studies on supercooled water, either $D\tau_{\alpha}$ or $D\tau_{\alpha}/T$
was tested for SE violation.  
However, the original SE relation $D\eta/T$ has not been widely studied
because of the high computational costs for calculating $\eta$,
particularly at low temperatures.
Therefore, to determine the origin of the SE violation, obtaining $\eta$ is important.
Hence, the central aims of the present study are
to obtain $\eta$ and to identify the time scales associated with $\eta$ and $D$
to reveal the origin of the SE violation in supercooled water.
 
The outline of the present study is as follows.
First, the SE violation in supercooled liquid water is examined using
molecular dynamics simulations of the
TIP4P/2005 model~\cite{Abascal:2005ka, Abascal:2010dw}.
In particular, comprehensive numerical calculations with respect to
shear viscosity are performed on the basis of the shear stress
correlation function, 
which are comparable with recent
studies for supercooled water using SPC/E (simple point
charge/extended)~\cite{Galamba:2017eq} and
TIP4P/2005f~\cite{Guillaud:2016bk}.
Our results provide a more systematic examination of the SE violation in
supercooled water.
The justification of the scenario $\eta/T\sim \tau_\alpha$ is
demonstrated, which is consistent with the previous
studies in simple liquids~\cite{Yamamoto:1998gb,
Sengupta:2013dg, Shi:2013ji, Kawasaki:2014ky}.

\begin{figure*}[t]
\centering
\includegraphics[width=.9\textwidth]{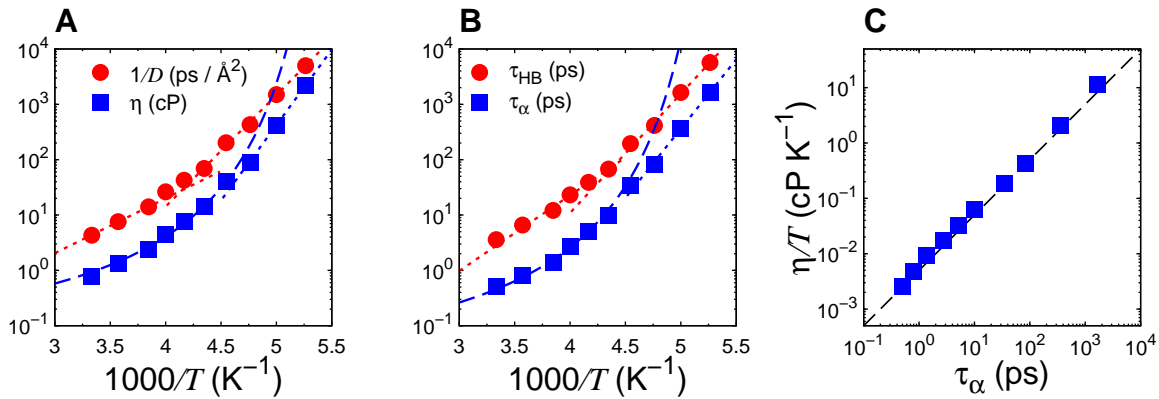}
\caption{\textbf{Dynamical properties in TIP4P/2005 supercooled water.}
(A) Temperature dependence of 
viscosity $\eta$ and 
translational diffusion constant $D$.
The blue dashed curve is the fitting of the Vogel-Fulcher-Tammann law $\eta\propto
 \exp(BT_0/(T-T_0))$ with $T_0=170$ K and $B=1.79$.
The blue dotted line is the Arrhenius law for $\eta\propto \exp(E_\mathrm{A}/T)$ at
 lower temperatures with an activation energy of $E_\mathrm{A}=52.1$ kJ/mol.
Arrhenius behaviors $D^{-1}\propto \exp(E_\mathrm{A}/T)$ in both the
 high and the low temperature ranges are also
 shown as two red dotted lines, with activation energies of $E_\mathrm{A}=19.0$ kJ/mol
 and $E_\mathrm{A}=38.6$ kJ/mol, respectively.
(B) Temperature dependence of the $\alpha$-relaxation time
 $\tau_\alpha$ and the HB lifetime $\tau_\mathrm{HB}$.
The blue dashed curve is the fitting of the Vogel-Fulcher-Tammann law $\tau_\alpha\propto
 \exp(BT_0/(T-T_0))$ with $T_0=175$ K and $B=1.87$.
The blue dotted line is the Arrhenius law for $\tau_\alpha\propto
 \exp(E_\mathrm{A}/T)$ at lower temperatures with an activation energy
 of $E_\mathrm{A} = 47.9$ kJ/mol.
Arrhenius behaviors $\tau_\mathrm{HB}\propto \exp(E_\mathrm{A}/T)$ in both
 the high and the low temperature ranges are also
 shown as two red dotted lines, with activation energies $E_\mathrm{A}=26.1$ kJ/mol
 and $E_\mathrm{A}=41.2$ kJ/mol, respectively.
 (C) Relationship between $\eta/T$ and $\tau_\alpha$.
The direct proportional relation $\eta/T \propto \tau_\alpha$ is obtained.
The dashed line is a guide to the eye.
}
\label{fig_tau}
\end{figure*}

Second, the role of the time scale associated with
hydrogen bond (HB) dynamics in the SE relation is investigated.
The rearrangement of the HB network in water
is expected to play a critical role in determining its
dynamical properties~\cite{Stillinger:1980ws, Ohmine:1993ij,
Luzar:1996gw, Starr:1999ki}.
In addition, the tetrahedrality due to the HB
network increases considerably
with decreasing temperature~\cite{Xu:2009hq, Kumar:2009dk, Saito:2013ff}.
This highly structured tetrahedral network is 
associated with the hypothesized liquid-liquid transition between
a high-density liquid and low-density liquid~\cite{Poole:1992ka,
Stanley:1998hf, Debenedetti:2003gn, Poole:2011bb, Kumar:2013fq,
Palmer:2014jb, Yagasaki:2014ew, Singh:2016bu},
although this scenario is currently controversial~\cite{Limmer:2011fu,
Limmer:2013iq, Overduin:2015tw}.
Thus, these facts necessitate
an investigation of the role of HB dynamics in the SE relation.
We show that the SE relation is preserved (SE preservation) when we use the
HB breakage time scales instead of $\tau_{\alpha}$, that is,
the strong coupling between the diffusion constant $D$ and HB lifetime $\tau_\mathrm{HB}$ at any temperature.
This preservation is attributed to the activated
jumps of mobile molecules that characterize the translational diffusion.

Third, the origin of the observed SE violation [that it, the decoupling between
the diffusion constant $D(\sim {\tau_\mathrm{HB}}^{-1})$ and $\alpha$-relaxation time $\tau_\alpha$]
is elucidated.
For this, non-Gaussian parameters and four-point dynamic correlations
are examined to probe the degree of dynamic heterogeneities in supercooled water.
Here, the SE violation/preservation is additionally demonstrated in terms of other significant time scales
such as the stress relaxation and the mobile/immobile contributions of the
dynamic heterogeneities.
From these classifications of various time scales, the degree of the SE violation is explained by the
increase in
the plateau modulus and the nonexponentiality of the stress correlation
function upon supercooling.
This elucidation for SE violation is also correlated with the growing of
non-Gaussianity and dynamic heterogeneities.

\begin{figure*}[t]
\centering
\includegraphics[width=.9\textwidth]{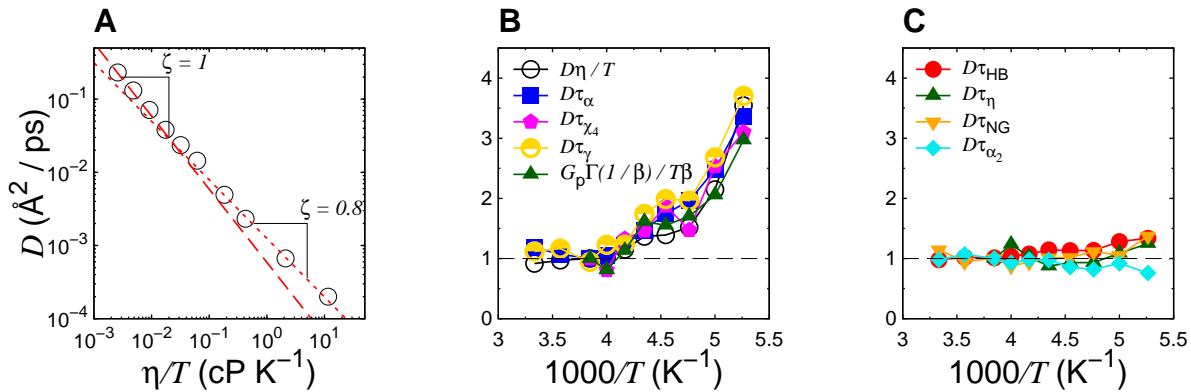}
\caption{\textbf{SE violation and preservation.}
(A) Translational diffusion constants $D$
vs. viscosity scaled by the temperature, $\eta /T$. 
Dashed and dotted lines present the fractional SE relation, $D \propto (\eta/T)^{-\zeta}$.
As $T$ decreases, the crossover of the power law exponent from
 $\zeta\approx 1.0$ (satisfying SE relation) to 0.8 (SE violation) is
 obtained by the fitting of the data.
(B) 
Inverse temperature dependence of the SE ratios: $D\eta/T$,
 $D\tau_{\alpha}$, $D\tau_{\chi_4}$, $D\tau_{\gamma}$, and $G_\mathrm{p} \Gamma(1/\beta)/(T\beta)$ scaled by
 their values at $T_\mathrm{A}=260$ K.  
All SE ratios exhibit the SE violation in the lower $T$ regime. 
Note that data of $D\tau_{\chi_4}$ and $G_\mathrm{p} \Gamma(1/\beta)/T\beta$ above $T_\mathrm{A}=260$ K is omitted.
(C) Inverse temperature dependence of the SE ratios: 
 $D\tau_\mathrm{HB}$, $D\tau_{\eta}$, $D\tau_\mathrm{NG}$, 
and $D\tau_{\alpha_2}$ scaled by their values at $T_\mathrm{A}=260$ K. 
All SE ratios satisfy the SE preservation, even at lower $T$.
Note that data of $D\tau_{\eta}$ above $T_\mathrm{A}=260$ K is omitted.
}
\label{fig_se}
\end{figure*}

\section*{Results}
\subsection*{SE violation}

The translational mean square displacement (MSD) was calculated at different
temperatures (see Materials and Methods).
The results are shown in fig.~S1A.
The diffusion constant was quantified from the long-time behavior of the
MSD (see Materials and Methods).
In Fig.~\ref{fig_tau}A, we plot the temperature dependence of $D$.
The overall behavior is in good agreement with the
previously reported result of the TIP4P/2005 model~\cite{Rozmanov:2012ja}.
The shear viscosity $\eta$ in the TIP4P/2005 supercooled water was
investigated from the stress correlation function $G_\eta(t)$ (see
Materials and Methods and fig.~S1B).
The shear viscosity $\eta$ was determined from Green--Kubo formula (see
Materials and Methods).
The temperature dependence of the viscosity $\eta$ is plotted in
Fig.~\ref{fig_tau}A along with that of $D$.
At $T=300$ K, the estimated value is $\eta\approx 0.78$ centipoise (cP), which is 
approximately the same as the reported value for
TIP4P/2005~\cite{Gonzalez:2010ez, GuevaraCarrion:2011ee, Tazi:2012fs,
Fanourgakis:2012ig}.
Furthermore, the structural relaxation of supercooled water is identified by the
incoherent intermediate scattering function $F_\mathrm{s}(k, t)$ (see
Materials and Methods).
The time evolution of $F_\mathrm{s}(k, t)$ at various temperatures is illustrated
in fig.~S1C.
As outlined in previous simulation~\cite{Gallo:1996hf, Sciortino:1996hk, Paschek:1999cp,
Gallo:2012cz, DeMarzio:2016hl},
the behavior of $F_\mathrm{s}(k, t)$ of supercooled water is characterized by
a two-step and nonexponential relaxation
below the onset temperature $T_\mathrm{A} \approx 260$ K.
Figure~\ref{fig_tau}B shows the temperature dependence of the
$\alpha$-relaxation time $\tau_\alpha$ (see the definition of
$\tau_\alpha$ in Materials and Methods).
In our calculations, the fragile-to-strong crossover (FSC) \textit{weakly} occurs at
approximately $T_\mathrm{L}\approx 220$ K.
Around this crossover temperature $T_\mathrm{L}$, the temperature dependence of
$\eta$ and $\tau_\alpha$ changes from non-Arrhenius to Arrhenius
behavior, as shown in Fig.~\ref{fig_tau}(A and B).
The FSC is expected as a sign of the compressibility maximum locus
(``Widom line'') originating from the liquid-liquid
transition~\cite{Liu:2005iy, Xu:2005ha}.
The observed $T_\mathrm{L}\approx 220$ K is in accord with the crossing
temperature at $1$ g cm$^{-3}$ of the Widom line determined in recent
TIP4P/2005 simulations~\cite{Sumi:2013fy, Russo:2014fl, Singh:2016bu}.

The relationship between $\eta/T$ and $D$ is presented in Fig.~\ref{fig_se}A.
The SE relation $D \sim (\eta /T)^{-1}$ holds at high $T$ but obeys
the fractional formula of SE relation $D \sim (\eta / T)^{-\zeta}$ with $\zeta 
\approx 0.8$ below $T_\mathrm{X}\approx 240$ K.
The crossover from $\zeta=1$ to $\zeta=0.8$ in the
fractional SE relation is similar to the recent
experimental result~\cite{Dehaoui:2015ii}.
This onset temperature appears to be above the FSC $T_\mathrm{L}\approx 220$ K.
As noted in Introduction, the alternative expressions for the SE
relation are conventionally examined via $D \sim {\tau_\alpha}^{-1}$ or
$D\sim(\tau_\alpha/T)^{-1}$.
The former formula uses the Gaussian approximation $F_\mathrm{s}(k, t)=\exp(-Dk^2 t)$.
If $\tau_\alpha$ is characterized by $\eta/T$, $D\tau_\alpha$ can play
the role of the SE relation.
Figure~\ref{fig_tau}C shows the proportional relationship
$\eta/T \sim \tau_\alpha$, which is
consistent with the previous results in simple liquids~\cite{Yamamoto:1998gb,
Sengupta:2013dg, Shi:2013ji, Kawasaki:2014ky}.
The temperature dependence of $D\tau_\alpha$ is
illustrated together with $D\eta/T$ in Fig.~\ref{fig_se}B.
This shows that $D\tau_\alpha$ is a good indicator of the SE violation
$D\eta/T$ below its onset temperature $T_\mathrm{X}\approx 240$ K.

\begin{figure}[t]
\centering
\includegraphics[width=.5\textwidth]{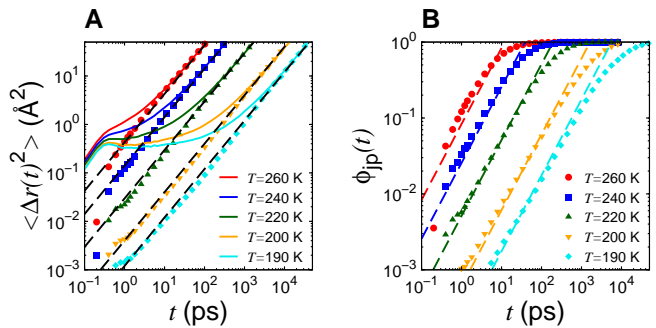}
\caption{\textbf{Diffusive properties of jump molecules.}
(A) Translational MSD $\langle \Delta r(t)^2\rangle$ for
 an O atom (solid curves), 
the MSD due to the \textbf{jp} O atoms 
$\langle \Delta {r_\mathrm{jp}}(t)^2 \rangle$ (points), and the Einstein
 relation $6D t$ (black dashed lines).
Here, $D$
is determined by the long-time asymptotic value of $\langle \Delta
 r(t)^2\rangle / 6t$ at each temperature.
For temperatures $T=190$, $200$, $220$, $240$, and $260$ K, 
$\ell_m$ is adjusted to $1.9$, $1.8$, $1.7$, $1.5$, and $1.4$ \AA, respectively.
(B) 
Average number fraction of the \textbf{jp} molecules $\phi_{\mathrm{jp}}(t)$.
Dashed lines represents the linear growth relations
 $t/\tau_\mathrm{HB}$ at each temperature.
}
\label{fig_jump_msd}
\end{figure}

\subsection*{SE preservation}

We introduce the \textit{generalized} SE ratio $D\tau$ with other
significant time scales in supercooled water.
First, we focus on the dynamics of HB breakage.
The number of the non-broken HBs for all molecules $N_\mathrm{HB}(t)$
was calculated in the time interval $t$, and then
the average number fraction $C_\mathrm{HB}(t)$ was calculated.
The results are shown in fig.~S1D [see the
detailed definitions of the HB and $C_\mathrm{HB}(t)$ in Materials and Methods].
The HB lifetime $\tau_\mathrm{HB}$ was then determined from
$C_\mathrm{HB}(t)$ (see the definition of $\tau_\mathrm{HB}$ in
Materials and Methods).
Its temperature dependence is displayed in 
Fig.~\ref{fig_tau}B along with that of $\tau_\alpha$.
Remarkably, both $D^{-1}$ and $\tau_\mathrm{HB}$
exhibit a \textit{similar} Arrhenius temperature dependence, which is
different from that of $\eta$ or $\tau_\alpha$ exhibiting the FSC.
Thus, we obtain a marked preservation of the SE relation (SE preservation),
$D \sim {\tau_\mathrm{HB}}^{-1}$ at any
temperature, as evident in Fig.~\ref{fig_se}C.
This SE preservation $D\sim {\tau_\mathrm{HB}}^{-1}$ implies that the appropriate time
scale associated with the translational diffusion
$D$ is not $\tau_\alpha$ but should instead be the HB lifetime $\tau_\mathrm{HB}$.
The HB breakage is commonly speculated to 
occur intermittently, inducing a markedly large number of jumping water molecules, particularly in supercooled states.
Exploring how the HB dynamics are
related to the translational diffusion via the jumping molecules is worthwhile.
This issue will be discussed later.

Next, we examine the stress relaxation time $\tau_{\eta}$.
The long-time behavior of $G_\eta(t)$ is well fitted by the
stretched-exponential function $G_\mathrm{p}\exp[-(t/\tau_\eta)^\beta]$
(see fig.~S1B).
$G_\mathrm{p}$ and $\tau_\eta$ 
denote the plateau modulus and the stress relaxation time, respectively.
The exponent $\beta(< 1)$ is the degree of nonexponentiality.
The temperature dependence of $G_\mathrm{p}$ and $\beta$ is illustrated
in fig.~S2A.
Note that the stress relaxation time $\tau_\eta$ 
differs from the relaxation time of the Maxwell model
$\tau_\mathrm{M} =  \eta/ G_{\infty}$ with the instantaneous shear modulus $G_\infty=G_\eta(t=0)$.
If the temperature dependence of $G_\infty$ is negligible, then the viscosity
$\eta$ is identified by $\tau_\mathrm{M}$.
Furthermore, provided that $\tau_\mathrm{M}$ equals $\tau_\alpha$, the linear relationship
$\eta \sim \tau_\alpha$ is obtained.
However, as seen in Fig.~S1B, $G_\infty$
and $G_\mathrm{p}$ increase slightly with decreasing temperature.
Instead of the Maxwell model, the viscosity $\eta$ is determined not
only by the stress relaxation
time $\tau_\eta$ but also by the plateau modulus $G_\mathrm{p}$.
This relationship will be clarified later.
We additionally obtained another preservation of the SE relation,
$D \sim \tau_{\eta}^{-1}$ at any temperature, as evident in Fig.~\ref{fig_se}C.
This observation implies that 
the HB breakage is correlated with the relaxation process of the local stress.

As mentioned in Introduction, 
the SE violation is possibly attributed to the heterogeneous dynamics,
that it,
coexistence of correlated mobile and immobile motions.
In this case, the distribution of the single-molecular displacement
becomes non-Gaussian at supercooled states.
When $F_\mathrm{s}(k, t)$ is described by the Gaussian approximation using the
MSD
$F_\mathrm{s}^{\mathrm{Gauss}}(k,t)=\exp[-k^2\langle\Delta r^2(t)\rangle/6]$, 
the relation $\tau_\alpha=(Dk^2)^{-1}$ at the diffusive regime is obtained~\cite{Hansen:2013uv}.
Therefore, the non-Gaussian behavior is directly linked with the SE violation.
Analogous to the previous study~\cite{Sciortino:1996hk}, 
the degree of the non-Gaussianity $\Delta F_\mathrm{s}(k, t)\equiv F_\mathrm{s}(k,
t)-F_\mathrm{s}^{\mathrm{Gauss}}(k,t)$
is plotted in fig.~S1C.
We introduce the peak time of $\Delta F_\mathrm{s}(k,t)$ as
$\tau_\mathrm{NG}$, which characterizes the time scale of the maximum
deviation from the Gaussian behavior.
As shown in Fig.~\ref{fig_se}C,
the ratio $D\tau_\mathrm{NG}$ represents the SE preservation at any temperature.
We also calculated the conventional non-Gaussian parameter
$\alpha_2(t)$ and determined
the peak time of $\alpha_2(t)$ as $\tau_{\alpha_2}$
[see the definition of $\alpha_2(t)$ in Materials and Methods and
fig.~S3A].
As demonstrated in Fig.~\ref{fig_se}C, the time scale $\tau_{\alpha_2}
(\simeq\tau_\mathrm{NG})$ is coupled with $D$ even at supercooled states.
From the definition, the first correction of cumulant expansion of
$\Delta F_\mathrm{s}(k, t)$ is given by $\alpha_2(t)$.
Thus, $\tau_\mathrm{NG}$ and
$\tau_{\alpha_2}$ exhibit similar temperature dependence.
A similar observation has been reported in Lennard--Jones
supercooled liquids~\cite{Starr:2013bs};
however, the SE ratio $D\tau_{\alpha_2}/T$ was used, contrary
to our results.

The relationship between the non-Gaussianity and the HB breakage is
discussed next.
The physical implication of SE preservation $D\tau_{\alpha_2}$ is also given.
Furthermore, effects of characteristic time scales of dynamic
heterogeneities on the SE
violation/preservation are examined.

\subsection*{Relationship between translational diffusion and HB breakage}

Let us examine how HB breakages are coupled with diffusion. 
To this end, we introduce the jumping($\textbf{jp}$) molecules with large displacements. 
Here, the \textbf{jp} molecules undergoing jumping motions are defined as those O
atoms that moved farther than an arbitrary cutoff length $\ell_m$, 
$\Delta r_i(t)= |\mathbf{r}_i(t)-\mathbf{r}_i(0)| > \ell_m$ during the time
interval $t$.
We calculate the MSD due to the \textbf{jp} O
atoms, $\langle\Delta {r_\mathrm{jp}}(t)^2 \rangle
=(1/N)\sum_{i\in{\mathrm{jp}}}\langle \Delta
r_i(t)^2\rangle$.
The summation is over the \textbf{jp} molecule number
$N_\mathrm{jp}(t)$ at time $t$.
In Fig.~\ref{fig_jump_msd}A, the \textbf{jp} component of the MSD, $\langle
\Delta r_\mathrm{jp}(t)^2\rangle$, is
plotted at several temperatures.
Because of the \textbf{jp} molecules, this restricted MSD exhibits
the diffusive behavior $6 D t$ even at short time regimes ($t \gs 1$ ps).
After a longer time, the \textbf{jp} contributions to the MSD
asymptotically reach the full MSD curves at each temperature because
all O atoms eventually move a distance greater than $\ell_m$.
In practice, the value of $\ell_m$ is adjusted to the long time regimes
of full MSD at each temperature.
For the temperature $T=190$ K, $\ell_m= 1.9$ \AA\ is chosen 
corresponding to the position at first shoulder of 
van Hove function
$G_\mathrm{s}(r,t)=\langle(1/N)\sum_{i=1}^N \delta(\mathbf{r}
-\mathbf{r}_i(t)+\mathbf{r}_i(0))\rangle$ with $r=|\mathbf{r}|$, 
which represents the distribution of single-molecular displacement
(see Fig.~S4, A to C).
At the time scale of $\tau_\mathrm{NG} \approx 1$ ns,
$G_\mathrm{s}(r, t)$ is largely deviated from the Gaussian form
$G_\mathrm{s}^{\mathrm{Gauss}}(r,
t)=[1/(4\pi Dt)^{3/2}]\exp(-r^2/4Dt)$.
This deviation
implies that the spatial distribution of single-molecular
displacement becomes heterogeneous.
In particular, a double-peaked structure for $G_\mathrm{s}(r, t)$ indicates
two distinct contributions due to jumping and nonjumping
molecules.
This non-Gaussianity can be clarified by
the decomposition of $G_\mathrm{s}(r, t)$ due to the
number of HBs broken $\mathcal{B}_i(t)$ during the time $t$ for the molecule $i$
[see the definition of $\mathcal{B}_i(t)$ in Materials and Methods].
The molecules having more than three broken HBs [$\mathcal{B}_i(t)>3$],
which destroy the molecules' local tetrahedral structures, are entirely
subjected to the jumping motions.
The displacements of these molecules exceed
the cutoff length $\ell_m=1.9$ \AA\ at 1 ns.
As demonstrated in the study by Kawasaki and Onuki~\cite{Kawasaki:2013bg}, 
this cutoff length $\ell_m$ enables the selection of irreversible jumps
as a result of an activation process analogous to nucleation~\cite{Onuki:2007up}.
The average number fraction of the \textbf{jp} molecules,
$\phi_\mathrm{jp}(t)\equiv \langle N_\mathrm{jp}(t)\rangle/N$,
exhibiting activation jumps increases linearly over time.
The jump rate is approximately given by
${\tau_\mathrm{HB}}^{-1}$, that is, $\phi_{\mathrm{jp}}(t) \simeq 
t/\tau_\mathrm{HB}$, which is demonstrated in Fig.~\ref{fig_jump_msd}B.
If the mean jump length $\ell_\mathrm{jp}$ is assumed, then
the \textbf{jp} component of the
MSD $\langle \Delta r_\mathrm{jp}(t)^2\rangle$ increases linearly with
time as $\ell_\mathrm{jp}^2 t/ \tau_\mathrm{HB}$ from short
time intervals.
As demonstrated in Fig.~\ref{fig_jump_msd}A, $\langle\Delta
r_\mathrm{jp}(t)^2\rangle$ exhibits $6Dt$.
Thus, these results clarify the correlation between translational
diffusion and HB breakage and agree with the 
demonstrated SE preservation, $D\sim {\tau_\mathrm{HB}}^{-1}$
(see again Fig.~\ref{fig_se}C).
Furthermore, the mean jump length can be estimated by
$\ell_\mathrm{jp}=\sqrt{6D\tau_\mathrm{HB}}\approx 2.6$ \AA.

\subsection*{Mechanism of SE violation}

The demonstrated SE violation indicates that the translational diffusion
constant $D$ is not
characterized by the $\alpha$-relaxation time $\tau_\alpha$.
The SE violation is explained in terms of the
peak height of $\Delta F_\mathrm{s}(k,t)$ at $\tau_\mathrm{NG}$, which is
represented by $\Delta F_\mathrm{s}^\mathrm{peak}$.
By using the SE preservation $D \sim {\tau_\mathrm{NG}}^{-1}$, we can express
the degree of SE violation in $D\tau_\alpha$ by the
temperature dependence of $\Delta
F^\mathrm{peak}_\mathrm{s}$ (see text S1 for details).
That is, the increase in the degree of non-Gaussianity 
is in accord with the degree of SE violation in $D\tau_{\alpha}$.

As mentioned above, the non-Gaussianity is directly relevant with
dynamic heterogeneities.
The observed double-peak structure of $G_s(r, t)$ at lower temperatures is the
main feature of dynamic heterogeneity (see fig.~S4, A to C).
Note that $\tau_\mathrm{\alpha_2}(\simeq \tau_\mathrm{NG})$
strongly characterizes the contribution of the mobile molecules that move
faster than the Gaussian distribution~\cite{Flenner:2005es}.
The peak time $\tau_{\alpha_2}$ of $\alpha_2(t)$ becomes smaller than the structural
relaxation time $\tau_\alpha$, particularly at low temperatures.
Up to the time scale $\tau_{\alpha_2}$, a tagged molecule is trapped by
the surrounding cage, which is observed as the plateau of MSD (see fig.~S1A).
The cage eventually breaks at $\tau_{\alpha_2}$ and then the tagged
molecule begins to
escape from the original position due to the jump motion.
This physical implication is consistent with the demonstrated SE
preservation $D\tau_\mathrm{HB}$.

The non-Gaussianity is additionally quantified by a new
non-Gaussian parameter $\gamma(t)$, which emphasizes the immobile and
slower contribution of dynamic heterogeneities [see the definition of
$\gamma(t)$ in Materials and Methods and fig.~S3B]~\cite{Flenner:2005es}.
The peak time $\tau_\gamma$ of $\gamma(t)$ becomes slower than
$\tau_{\alpha_2}$ with decreasing temperature.
This indicates 
the decoupling between mobile and immobile molecules in supercooled states.
As demonstrated in Fig.~\ref{fig_se}B, 
the SE ratio $D\tau_\gamma$ exhibits the SE violation, following the
similar temperature dependence of $D\tau_\alpha$.
Another quantity to examine the dynamic heterogeneities is the
four-point correlation function $\chi_4(k, t)$ that is defined by the
variance of $F_s(k, t)$ [see the definition of
$\chi_4(k, t)$ in the Methods section and fig.~S3C]~\cite{Toninelli:2005ci}.
The value of $\chi_4(k, t)$ is related to the correlation length of
dynamic heterogeneities at the time scale $t$.
As demonstrated in Fig.~S3C,
$\chi_4(k, t)$ exhibits the peak value at $\tau_\alpha$,
which increases as the temperature decreases.
Figure~\ref{fig_se}B shows that
the peak time $\tau_{\chi_4}$ of
$\chi_4(k, t)$ also acts as the SE violation.
These results indicate that the immobile and slower component of non-Gaussianity
is characterized by the time scales $\tau_\alpha$ and
$\tau_\gamma$ presenting the SE violation.
In contrast, 
the time scales $\tau_\mathrm{NG}$, 
$\tau_\mathrm{\alpha_2}$, and $\tau_\mathrm{HB}$ is coupled with the 
diffusion constant $D$, which is markedly governed by the mobile and jumping
molecules.

Furthermore, the increase in the degree of the
non-Gaussianity $\Delta F^\mathrm{peak}_\mathrm{s}$ upon supercooling
can be interpreted by the viscoelasticity and nonexponentiality in the stress relaxation
function $G_\eta(t)$.
The viscosity $\eta$ is mainly determined by $G_\mathrm{p}$ and $\tau_\eta$
according to the long-time behavior of $G_\eta(t)\simeq
G_\mathrm{p}\exp[-(t/\tau_\eta)^\beta]$ (see fig.~S1B).
This dependence of $G_\eta$ on $G_\mathrm{p}$ and $\tau_\eta$ leads to
the approximation of $\eta$ as $\int_0^\infty
G_\mathrm{p}\exp[-(t/\tau_\eta)^\beta]
dt=G_\mathrm{p}\tau_\eta\Gamma(1/\beta)/\beta$, where $\Gamma(\cdots)$
is the gamma function.
Figure~S2A shows that the plateau modulus
$G_\mathrm{p}$ increases, whereas the stretched exponent $\beta$ decreases
with decreasing temperature.
The clear correlation between $\eta$ and
$G_\mathrm{p}\tau_\eta\Gamma(1/\beta)/\beta$ is demonstrated in
fig.~S2B except for high temperatures.
The plateau moduli are well developed below
$T_\mathrm{A}\approx 260$ K, which is correlated with the onset of the
two-step relaxation in $F_\mathrm{s}(k, t)$.
By combining it with $D\sim {\tau_\mathrm{HB}}^{-1}$,
we obtain the relationship $D\eta/T \sim (G_\mathrm{p}\Gamma(1/\beta)/T\beta)\times
(\tau_\eta/\tau_\mathrm{HB})$.
The linear relationship between $\tau_\mathrm{HB}$ and $\tau_\eta$
provides an alternative representation for SE violation as $D\eta/T
\sim G_\mathrm{p}\Gamma(1/\beta)/T\beta$, as demonstrated in Fig.~\ref{fig_se}B.
Additionally, 
the SE violation is attributed to the immobile 
molecules within dynamic heterogeneities, whose time scales are
$\tau_\alpha$, $\tau_\gamma$, and $\tau_{\chi_4}$.
This decoupling is in accord with the development
of $G_\mathrm{p}$, that is, the emergence of solid like regions.
Therefore, the increase in the non-Gaussianity $\Delta
F^\mathrm{peak}_\mathrm{s}$ is 
directly relevant to the increase in
$G_\mathrm{p}$ (attained solidity) and to the decrease in 
$\beta$ (increase in the nonexponentiality for the stress relaxation), 
resulting in the SE violation with lowering $T$.

\section*{Discussion}
In summary, we reported comprehensive numerical results concerning
the SE relation in the TIP4P/2005 supercooled water.
In particular, the temperature dependence of the shear viscosity was
quantified from the stress correlation function in a wide temperature
range (190 to 300 K).
Thus, the SE relation in supercooled liquid water was systematically examined as follows.

We reported that
the violation of the SE relation is characterized by the fractional form, $D \sim (\eta/T)^{-\zeta}$ 
with $\zeta \approx 0.8$.
The onset temperature of SE violation $T_\mathrm{X}\approx 240$ K is
slightly below $T_\mathrm{A}\approx 260$ K, which is the onset temperature of
the two-step relaxations exhibited in $F_\mathrm{s}(k, t)$ and $G_\eta(t)$.
These temperatures are above 
the FSC temperature
$T_\mathrm{L} \approx 220$ K observed in the temperature dependence of $\eta$ and
$\tau_\alpha$.
A similar observation, $T_\mathrm{L}< T_\mathrm{X} \ls T_\mathrm{A}$,
has been reported in numerical results using ST2 water model~\cite{Poole:2011bb}.
Furthermore, the degree of the SE violation was identified by $D\tau_\alpha$
from the proportional relation $\eta/T \sim \tau_\alpha$.
We also explored the role of HB breakage on the SE relation.
The results revealed that the time scale associated with
the translational diffusion constant $D$ should be 
the HB lifetime $\tau_\mathrm{HB}$, in accordance with the
preservation of the SE relation $D\sim {\tau_\mathrm{HB}}^{-1}$ even for
supercooled states. 
We observed that both $D$ and $\tau_\mathrm{HB}$ exhibit an
Arrhenius temperature dependence with a similar activation energy.
This SE preservation proposes the temperature independent length scale
$\ell_\mathrm{jp}=\sqrt{6D\tau_\mathrm{HB}}\approx 2.6$ \AA, which has no
relation with
the Widom line and the possible liquid-liquid transition.

We quantitatively confirmed that the observed preservation of the SE relation $D\sim
{\tau_\mathrm{HB}}^{-1}$ was attributed to
the effect of the activated jumping of mobile molecules on the translational
diffusion.
The distinction between jumping and nonjumping molecules in supercooled states is a manifestation
of spatially heterogeneous dynamics, that is, the dynamic
heterogeneities in supercooled water~\cite{Giovambattista:2004ft, Giovambattista:2005ib}.
In particular, the MSD from the \textbf{jp}
molecules, $\langle \Delta r_\mathrm{jp}(t)^2\rangle$, was characterized by the diffusive
behavior $6Dt$, even on short time scales.
The jumping rate was characterized by the inverse of the HB lifetime $\tau_\mathrm{HB}$.
An analogous result showing the SE preservation between $D$ and $\tau_\mathrm{HB}$ has
already been obtained in both binary
soft-sphere mixtures (fragile liquids)~\cite{Kawasaki:2013bg}
and silica-like network-forming liquids (strong liquids)~\cite{Kawasaki:2014ky}.
In these studies, the bond-breakage method characterizing the changes in local
particle connectivity was used, which is essentially the same as the current
analysis regarding the HB network in liquid water.

Furthermore, we categorized other time scales (such as stress relaxation time $\tau_\eta$, 
time scales of the non-Gaussianity $\tau_{\alpha_2}$, $\tau_\gamma$, and
$\tau_\mathrm{NG}$, and four point dynamic susceptibility
$\tau_{\chi_4}$) into the SE violation and preservation.
Here, the time scales of $\tau_\eta$, $\tau_{\alpha_2}$, and $\tau_\mathrm{NG}$
characterize the mobile molecules within dynamic heterogeneities and are
coupled with the diffusion constant $D$ even for supercooled states.
In contrast, $\tau_\gamma$ and $\tau_{\chi_4}$ 
exhibit the temperature dependence similar to that of the
$\alpha$-relaxation time $\tau_\alpha$.
These time scales are governed by the immobile and slower molecules
and are decoupled with $D$ when the temperature decreases,
leading to the SE violation.

Finally, we revealed that the SE violation was
attributed to the increase in the degree of the non-Gaussianity $\Delta
F^\mathrm{peak}_\mathrm{s}$. 
Simultaneously, the SE relation is represented by
$D \eta/T \sim G_\mathrm{p}\Gamma(1/\beta)/T\beta$, where $G_\mathrm{p}$
and $\beta$ denote the plateau modulus and the stretched exponent in the
stress relaxation function, respectively.
Here, the proportional relationship between the stress relaxation time and
the HB lifetime $\tau_{\eta} \sim \tau_\mathrm{HB}$ was used.
Therefore, the time scales supporting the violation or preservation of
the SE relation were thoroughly identified;
attained solidity (increasing $G_\mathrm{p}$) and increasing nonexponentiality (decreasing $\beta$) 
give rise to the SE violation with decreasing the temperature.
Note that the nonexponentiality in the stress relaxation
is also a significant hallmark of the dynamic heterogeneities~\cite{Furukawa:2011bg}.
In our simulations, the plateau modulus and the nonexponentiality
develop largely below $T_\mathrm{A} \approx 260$ K.
Correspondingly, the growths of the non-Gaussianity and the dynamic
susceptibility are noticeable, as demonstrated in figs.~S3 (A to C).

There are other implications in developing the plateau modulus $G_\mathrm{p}$.
The SE violation with decreasing temperature will be relevant with
the decoupling between translational and rotational
motions in supercooled water.
It is expected that translational relaxations strongly
become slower, whereas molecules undergo rotational motions even inside
immobile solid like regions~\cite{Mazza:2007kr}.
The mechanism of this decoupling will be clarified in terms of the
attained solidity $G_\mathrm{p}$.
In addition, a recent theoretical study has shown that
the spatially heterogeneous dynamics is attributed to the thermal excitation
between the different metabasins of the free energy landscape~\cite{Yoshino:2012hj}.
In the framework, 
the value of the plateau modulus $G_\mathrm{p}$ is determined 
by the curvature of the local metabasin.
Considering these investigations,
the demonstrated SE preservation $D\tau_\mathrm{HB}$
will provide deeper insight into
the activated jump events occurring between different metabasins,
not only in supercooled water but
also in various glassy systems, although further investigations are required
to confirm it.

\section*{Materials and Methods}
\subsection*{Simulations}
The molecular dynamics simulations 
of liquid water were performed using the LAMMPS package~\cite{Plimpton:1995wl}.
The TIP4P/2005 model was used for the water
molecules~\cite{Abascal:2005ka, Abascal:2010dw}.
The NVT ensemble for $N=1000$ water molecules was first simulated at
various temperatures ($T=300$, 280, 260, 250, 240, 230, 220, 210, 200 and
190 K) with a fixed density $\rho=1$ g cm$^{-3}$
The corresponding linear dimension of the system is $L=31.04$ \AA.
After equilibration for a sufficient time at each temperature, the NVE ensemble simulations
were completed, yielding five independent 100 ns trajectories from which the various physical
quantities were calculated.
The simulations were performed with a time step of 1 fs.
The total CPU (central processing unit) time approximated about 20 years of single core time.

\subsection*{Incoherent intermediate scattering function and MSD}
The
incoherent intermediate scattering function is given by
\begin{equation}
F_\mathrm{s}(k, t)=\left\langle\frac{1}{N}
\sum_{i=1}^N\exp[i\mathbf{k}\cdot(\mathbf{r}_i(t)-\mathbf{r}_i(0))]\right\rangle,
\end{equation}
where $\mathbf{r}_i(t)$ is the position vector of the O atom of 
the water molecule $i$ at time $t$.
The bracket indicates an average over the initial time $t=0$.
The wave number $k=|\mathbf{k}|$ was chosen as $k=3.0$
\AA ${}^{-1}$, which corresponds to the first
peak position of the static structure factors of the O atom.
The $\alpha$-relaxation time $\tau_\alpha$ was determined by the fitting $F_\mathrm{s}(k, t)$
with 
$(1-f_\mathrm{c}) \exp[-(t/\tau_\mathrm{s})^2] + f_\mathrm{c}
\exp[-(t/\tau_\alpha)^{\beta_\alpha}]$, where $f_\mathrm{c}$, $\tau_\mathrm{s}$,
$\tau_\alpha$, and $\beta_\alpha$ are fitting parameters.
The exponent $\beta_{\alpha}$ is the degree of nonexponentiality of
$F_\mathrm{s}(k, t)$.

The MSD of
the O atom,
\begin{equation}
\langle \Delta r(t)^2\rangle = \left\langle\frac{1}{N}
\sum_{i=1}^N|\mathbf{r}_i(t)-\mathbf{r}_i(0)|^2\right\rangle ,
\end{equation}
was also calculated.
The translational diffusion
constant $D$ was determined from the long-time behavior of the MSD using
the Einstein relation, $D=\lim_{t\to\infty}\langle \Delta
r(t)^2\rangle / 6t$.

\subsection*{HB breakage and its lifetime}
The dynamics of HB was investigated by using r-definition~\cite{Kumar:2007bs},
where only the intermolecular O-H distance $r_\mathrm{OH}$ is involved.
An HB bond is present at the initial time if the $r_\mathrm{OH}$ is
less than $2.4$ \AA, corresponding to the first minimum of the
radial distribution function $g_\mathrm{OH}(r)$.
At a later time $t$, the HB is broken when the distance $r_\mathrm{OH}$
becomes larger than $2.4$ \AA, which is determined from the second minimum position of
$g_\mathrm{OH}$(r).

First, to characterize the local configuration change, we defined the number
of HBs broken during time $t$ for molecule $i$ as $\mathcal{B}_i(t)$.
Next, the characteristic time scale (that is, the HB lifetime $\tau_\mathrm{HB}$) was determined.
The number of HBs was calculated at the initial time $0$ and denoted as
$N_\mathrm{HB}(0)$. 
At time $t$, the number
of remaining HBs, $N_\mathrm{HB}(t)=N(0)-\sum_{i} \mathcal{B}_i(t)/2$, was less than the initial value
$N_\mathrm{HB}(0)$ due to HB breakages~\cite{Luzar:1996gw, Starr:1999ki}.
The average fraction of HB bonds as a function of time $t$ was then
defined as
\begin{equation}
C_\mathrm{HB}(t) = \langle N_\mathrm{HB}(t)/N_\mathrm{HB}(0) \rangle.
\end{equation}
The average HB lifetime $\tau_\mathrm{HB}$ was determined by fitting $C_\mathrm{HB}(t)$
with $\exp[-(t/\tau_\mathrm{HB})^{\beta_\mathrm{HB}}]$,
where the exponent $\beta_\mathrm{HB}$ is the degree of nonexponentiality of $C_\mathrm{HB}(t)$.

Furthermore, the present scheme 
is identical to the
bond-breakage method applied to various supercooled
liquids~\cite{Yamamoto:1997gu, Yamamoto:1998jg, Shiba:2012hm,
Kawasaki:2013bg, Kawasaki:2014ky, Shiba:2016bi}.
These previous studies have demonstrated that
the bond-breakage method is more remarkable when the
collective motions and dynamic heterogeneities peculiar to supercooled
states are characterized.

\subsection*{Stress correlation function and shear viscosity}

The autocorrelation function of the off-diagonal stress tensor is given by
\begin{equation}
G_{\alpha\beta}(t) = \frac{V}{k_\mathrm{B}T} \langle \sigma_{\alpha\beta}(t)
 \sigma_{\alpha\beta}(0)\rangle,
\end{equation}
where $V$ is the volume of the system and 
$\sigma_{\alpha\beta}$ represents the $\alpha\beta(\alpha \mathrm{and}
\beta=x, y, z)$ components of the
off-diagonal stress tensor and $k_\mathrm{B}$ is the Boltzmann constant.
The average stress correlation function is defined as
$G_\eta(t) = [G_{xy}(t)+G_{xz}(t)+G_{yz}(t)]/3$.
The shear viscosity $\eta$ was determined from the integral of
$G_\eta(t)$ as
$
\eta = \int_0^\infty G_\eta (t) dt,
$
using the Green--Kubo formula.

\subsection*{Characterizations of dynamic heterogeneities}

The non-Gaussian parameter for displacements of the molecules is the
conventional quantity to characterize dynamic heterogeneities in
various glass-forming liquids.
The equation is given by 
\begin{equation}
\alpha_2(t) =  \frac{3}{5}\frac{\langle \Delta r(t)^4\rangle}{\langle
 \Delta r(t)^2 \rangle^2}-1, 
\end{equation}
which represents the degree of the deviation 
from the Gaussian approximation in the density correlation function, which
is revealed by the cumulant expansion such as
\begin{equation}
F_\mathrm{s}(k,t) \sim
 F_\mathrm{s}^\mathrm{Gauss}(k,t)\left\{1+\frac{1}{2!}\alpha_2(t)[k^2\langle
				  \Delta r(t)^2\rangle/6]^2 \right\},
\end{equation}
where $F_\mathrm{s}^\mathrm{Gauss}(k,t) = \exp{(-k^2\langle \Delta r(t)^2\rangle/6)}$.
The difference is then given by
$\Delta F_\mathrm{s}(k,t) = F_\mathrm{s}(k,t)
-F_\mathrm{s}^\mathrm{Gauss}(k,t)$.

This $\alpha_2(t)$ is mainly dominated by mobile components in the
distribution of single-molecular displacement $G_s(r, t)$.
To emphasize immobile and slower components, 
another type of non-Gaussian parameter is given by
\begin{equation}
\gamma(t) =  \frac{1}{3}{\langle \Delta r(t)^2\rangle}\left\langle
 \frac{1}{\Delta r(t)^2} \right\rangle-1, 
\end{equation}
which is referred to as new non-Gaussian parameter~\cite{Flenner:2005es}.

Furthermore, the four-point dynamic susceptibility $\chi_4(k, t)$ is used
to identify the magnitude of dynamic heterogeneities.
The equation is defined from the variance of $F_s(k, t)$,
\begin{equation}
\chi_4(k, t) =  N\left\langle \frac{1}{N} \sum_{i=1}^N \left[\delta
						    F_i(\mathbf{k}, t)\right]^2
\right\rangle,
\end{equation}
where $
\delta F_i(\mathbf{k}, t) =  
\cos\{\mathbf{k}\cdot[\mathbf{r}_i(t)-\mathbf{r}_i(0))]\} - F_\mathrm{s}(k, t) $
is the $i$th molecular fluctuation in the real-part of density
correlator~\cite{Toninelli:2005ci}.

\begin{acknowledgements}
We thank A. Onuki for fruitful discussions.
K.K. is grateful to T. Nakamura for helpful information on computation.
This work was partially supported by JSPS
KAKENHI Grants No. JP15H06263 and No. JP16H06018 (T.K.), and
 No. JP26400428 and No. JP16H00829 (K.K.).
The numerical calculations were performed at Research Center of Computational
Science, Okazaki, Japan.
\end{acknowledgements}

\widetext
\clearpage

\begin{center}
\textbf{\large Supplementary Information}
\end{center}

\setcounter{equation}{0}
\setcounter{figure}{0}
\setcounter{table}{0}
\setcounter{page}{1}
\makeatletter
\renewcommand{\theequation}{S\arabic{equation}}
\renewcommand{\figurename}{fig.}
\renewcommand{\thefigure}{S\arabic{figure}}

\begin{figure}[H]
\centering
\includegraphics[width=.6\textwidth]{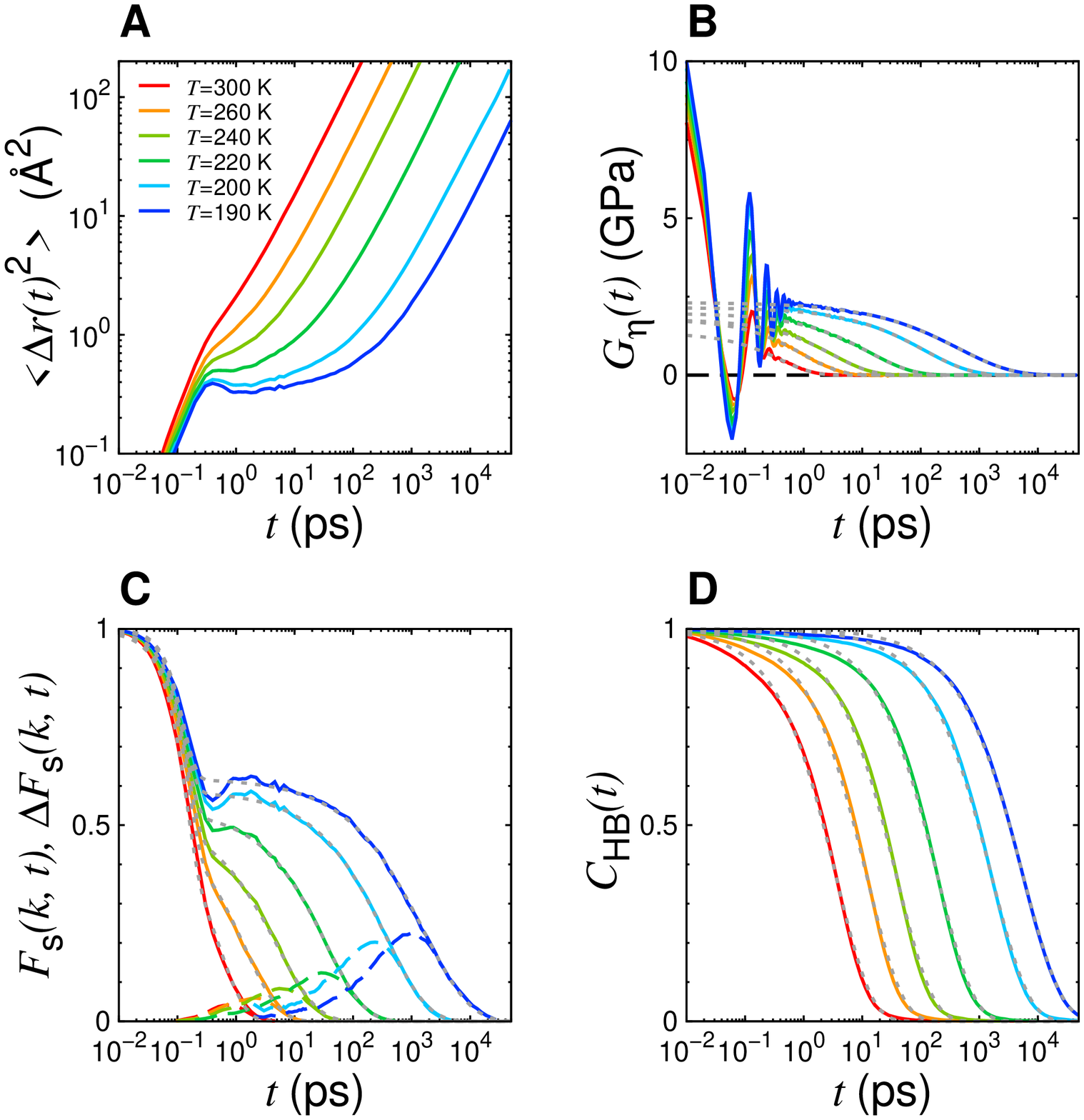}
\caption{\textbf{Time correlation functions.}
(A) Translational MSD $\langle \Delta r(t)^2\rangle$ for an O atom.
(B) Stress correlation function $G_\eta(t)$.
Long-time behaviors of $G_\eta(t)$ can be fitted by 
stretched-exponential fittings $G_\mathrm{p}\exp[-(t/\tau_\eta)^\beta]$
 (gray dashed curves).
(C) Incoherent intermediate scattering function $F_\mathrm{s}(k, t)$ for an O atom
as well as its deviation from the Gaussian approximation $\Delta F_\mathrm{s}(k, t) = F_\mathrm{s}(k,
 t)-F_\mathrm{s}^{\mathrm{Gauss}}(k, t)$ (dashed curves).
Gray dashed curves represent 
the fitting results using the function $F_\mathrm{s}(k, t)= (1-f_\mathrm{c})
 \exp[-(t/\tau_\mathrm{s})^2] + f_\mathrm{c} \exp[-(t/\tau_\alpha)^{\beta_\alpha}]$.
Wave number $k$ was chosen as $k=$3.0 \AA${}^{-1}$.
(D) Average number fraction of HB number $C_\mathrm{HB}(t)$.
Gray dashed curves represent 
the fitting results using the function $C_\mathrm{HB}(t)= 
 \exp[-(t/\tau_\mathrm{HB})^{\beta_\mathrm{HB}}]$.
}
\label{fig_tcf}
\end{figure}

\begin{figure}[H]
\centering
\includegraphics[width=.6\textwidth]{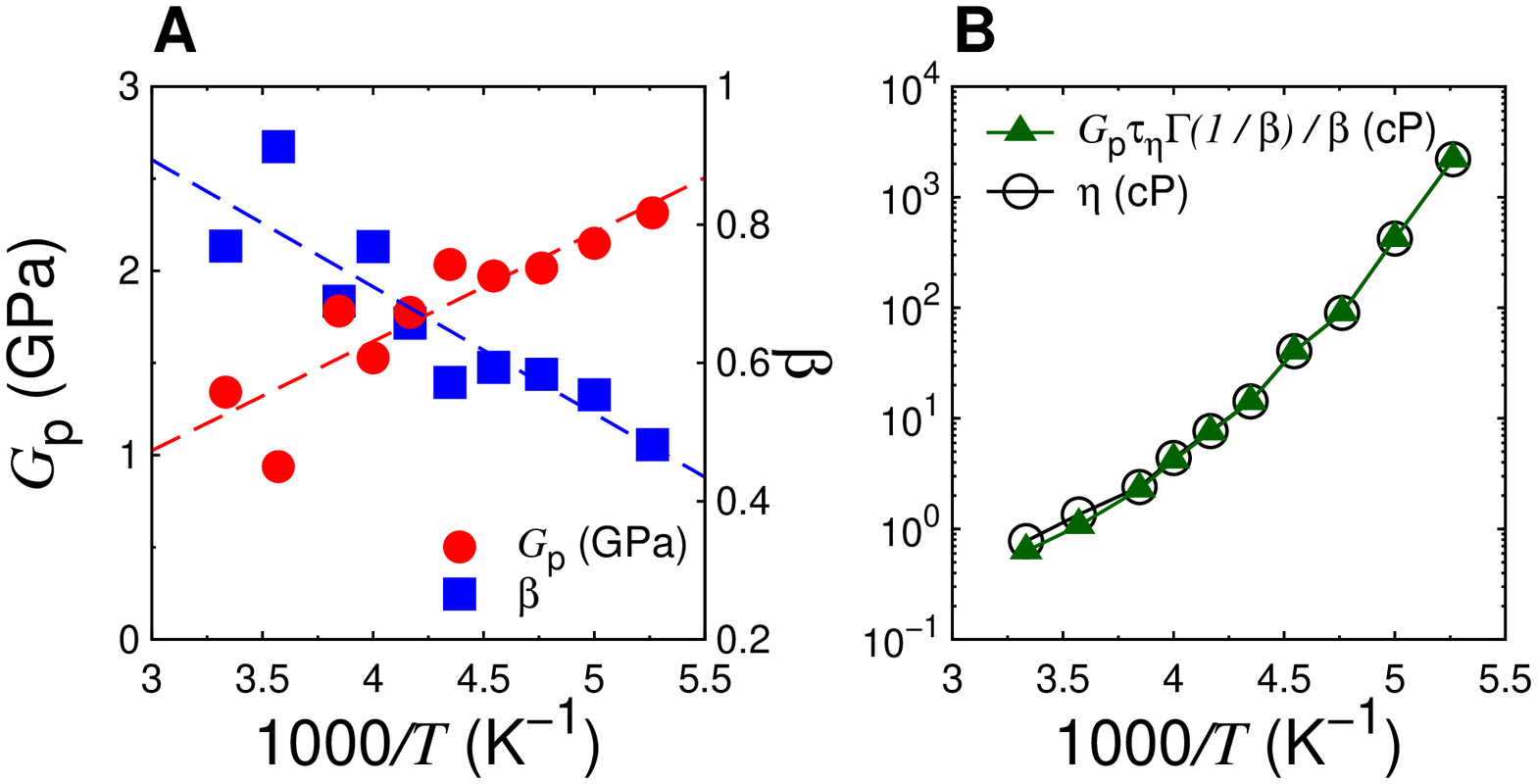}
\caption{\textbf{Properties of stress-relaxation function.}
(A) Temperature dependence of plateau modulus $G_\mathrm{p}$ (left axis)
 and stretched exponent $\beta$ (right axis).
The dashed lines are guides to the eye.
(B) Temperature dependence of $\eta$ and $G_\mathrm{p}\tau_\eta\Gamma(1/\beta)/\beta$.
}
\label{fig_tau_eta}
\end{figure}

\begin{figure}[H]
\centering
\includegraphics[width=.9\textwidth]{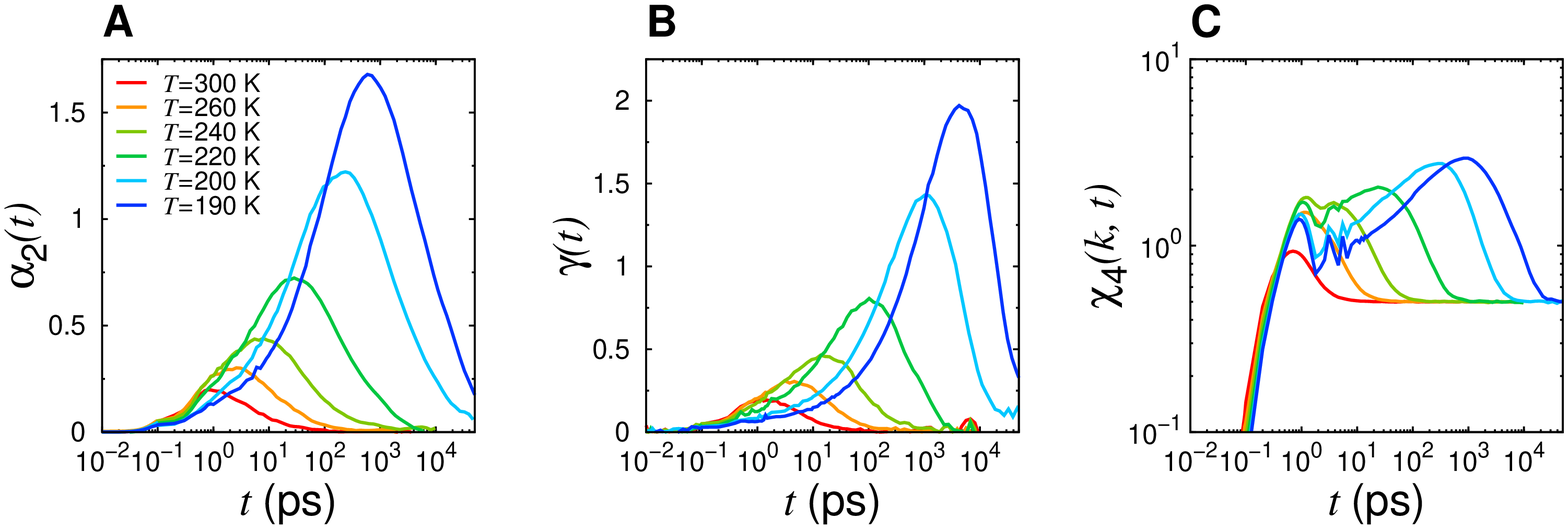}
\caption{\textbf{Characterizations of dynamic heterogeneities.}
(A) Non-Gaussian parameter $\alpha_2(t)$.
(B) New non-Gaussian parameter $\gamma(t)$.
(C) Four-point dynamical susceptibility $\chi_4(k, t)$.
Wave number $k$ was chosen as $k=$3.0 \AA${}^{-1}$.
}
\label{fig_dh}
\end{figure}

\begin{figure}[H]
\centering
\includegraphics[width=.9\textwidth]{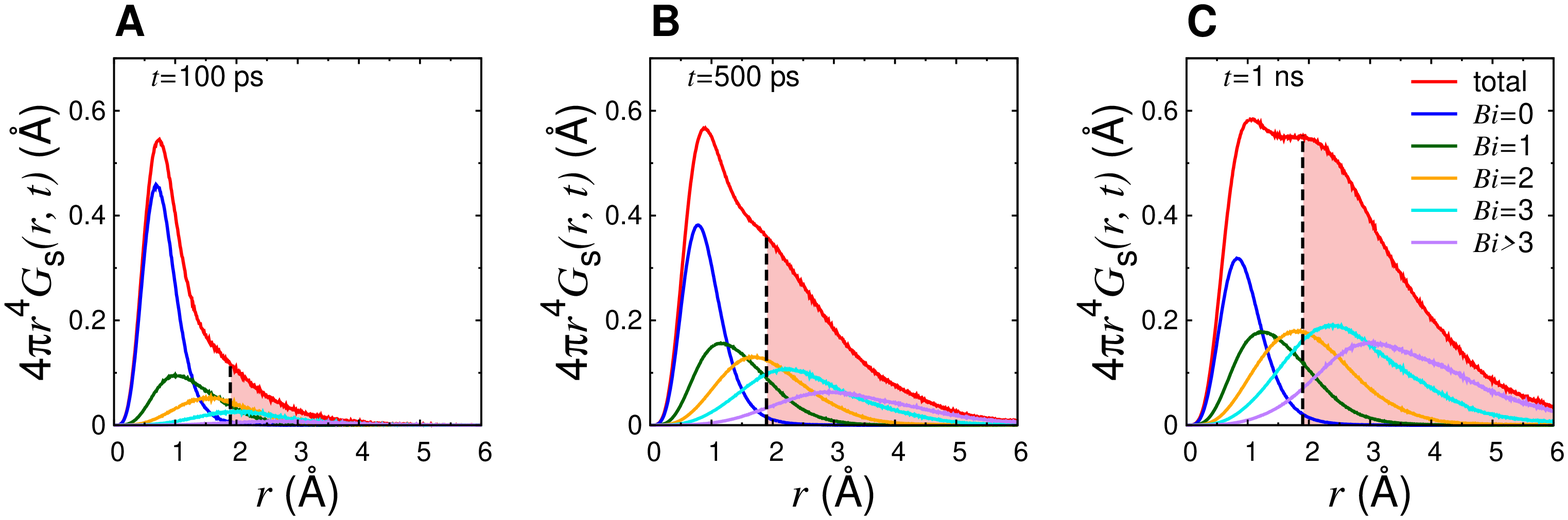}
\caption{\textbf{van Hove correlation functions.}
(A) Self-part of van Hove function $G_\mathrm{s}(r, t)$ at $t=100$ ps, 500 ps, and  1 ns
for $T=190$ K. 
The colored area represents the mobile component of the MSD due to the
 \textbf{jp} O atoms for which the displacement
exceeds jump length $\ell_m=1.9$ \AA\ during the time interval $t$, which is described
by $\langle \Delta {r_\mathrm{jp}}(t)^2 \rangle= \int_{\ell_m}^\infty 4\pi r^4 G_\mathrm{s}(r, t) dr$.
$G_\mathrm{s}(r, t)$ components arising from number of broken HB $\mathcal{B}_i(t)$
for molecule $i$ are also plotted.}
\label{fig_gs}
\end{figure}

\subsection*{text S1. SE violation evaluated by non-Gaussianity.}
Here, we show that the SE violation represented by $D\tau_{\alpha}$ can be evaluated 
by the peak height of non-Gaussianity for the intermediate scattering function:
$\Delta F^\mathrm{peak}_\mathrm{s}=\Delta F_\mathrm{s}(k,\tau_\mathrm{NG})$ (see fig.~S1C).
First, the function is written as
\begin{equation}
\Delta
 F^\mathrm{peak}_\mathrm{s}(k,t_\mathrm{NG})=F_\mathrm{s}(k,\tau_\mathrm{NG})-F_\mathrm{s}^\mathrm{Gauss}(k,\tau_\mathrm{NG})
\simeq  f_\mathrm{c}
\exp{\left\{-\left(\frac{\tau_\mathrm{NG}}{\tau_{\alpha}}\right)^{\beta_\alpha}
     \right\}}-\exp{(-k^2D\tau_\mathrm{NG})}.
\label{delta_F}
\end{equation}
When $T$ changes, SE preservation $D\tau_\mathrm{NG} \equiv C_1$
(constant) (see Fig.~2C) and  $\exp{(-k^2D\tau_\mathrm{NG})}
\equiv C_2$ (constant) are obtained. 
Therefore, by eliminating $\tau_\mathrm{NG}$ from the Eq. (\ref{delta_F}) we find
\begin{equation}
D\tau_{\alpha} = \frac{1}{C_1}\left\{\log\frac{f_\mathrm{c}}{\Delta
			       F^\mathrm{peak}_\mathrm{s}+C_2}
			      \right\}^{-1/\beta_\alpha} \propto \left\{
			      \log\frac{f_\mathrm{c}}{\Delta
			      F^\mathrm{peak}_\mathrm{s}}
								 \right\}^{-1/\beta_\alpha},
\label{se_nongauss}
\end{equation}
where $C_2\sim 0$ in the low $T$ regime.
This equation reveals that $D\tau_{\alpha}$ is increased when $\Delta
F^\mathrm{peak}_\mathrm{s}/f_\mathrm{c}$ is increased by lowering $T$.
Therefore, the SE violation is attributed to the increase in the degree of non-Gaussianity.
\end{document}